\documentclass{article}
\usepackage[numbers,sort&compress]{natbib}
\usepackage{spconf,amsmath,graphicx,hyperref}
\usepackage{color}
\usepackage{multirow} 
\usepackage{bm}



\title{A User-Item Aware Encoding Framework for Short Video}
\name{Weihui Deng, He Liu, Bin Wang, Xiaocheng Li, Deliang Fu,  Zhen Wang}
\address{Bytedance, Beijing, China}

\begin{document}
%
\maketitle
\vspace{-20pt}

\begin{abstract}
Traditional encoding methods from the Professional Generated Content (PGC) era face significant challenges due to two unique characteristics of user-generated short videos: massive daily uploads (reaching hundreds of millions) and heterogeneous content-consumer relationships (varying video quality and diverse consumer contexts). To address these challenges, we propose UIAE (User-Item Aware Encoding), a novel multiple bitrate ladder group encoding method comprising three key components: 1) Establishing user-item relationships via rule-based or DNN-based models; 2) Developing local optimization models maximizing Quality of Experience (QoE) for sub-populations based on contextual consumption patterns; 3) Deriving globally optimal encoding strategies through hierarchical model integration. Experimental results demonstrate that the UIAE framework  adaptively adjusts encoding outcomes for distinct user groups with varying context profiles, ensuring optimal QoE for end viewers. We have deployed the UIAE method into the online pipeline, handling tens of millions of video transcoding tasks daily.
\end{abstract}

\vspace{-5pt}
\section{Introduction}
\vspace{-5pt}

\begin{figure}  
  \centering
  \includegraphics[width=0.9\columnwidth]{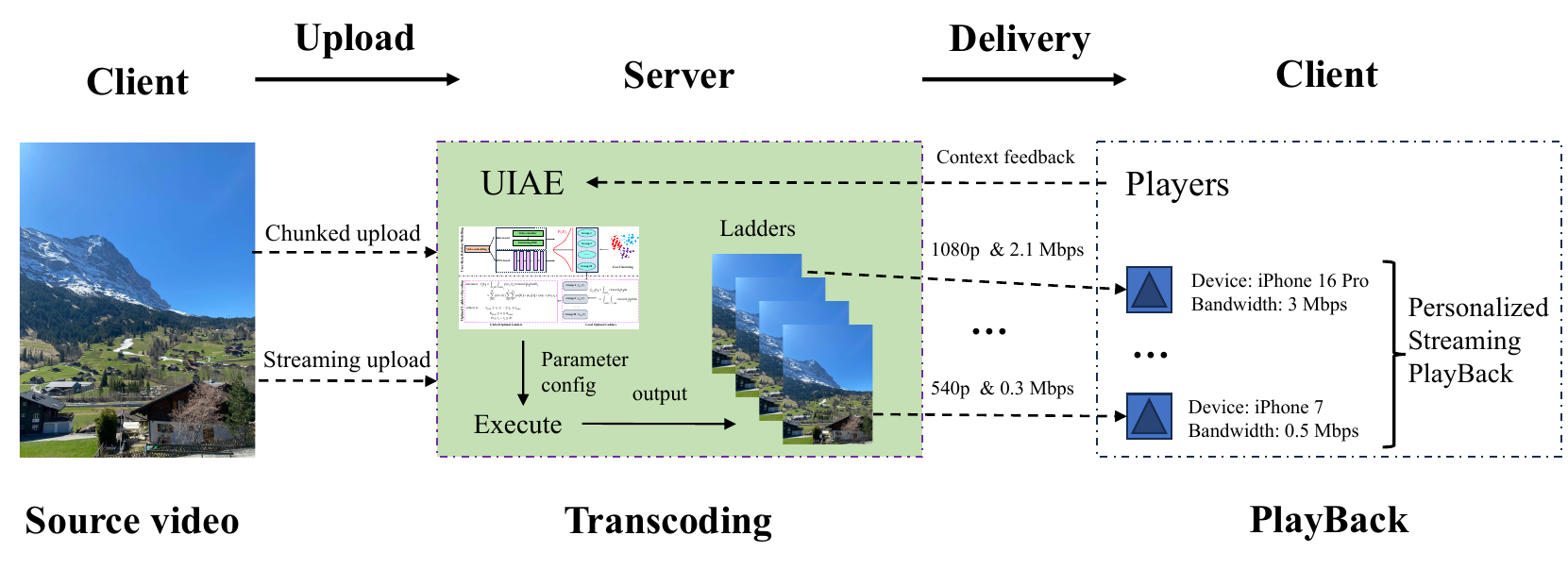}
  \caption{The End-to-End Audiovisual Pipeline.}
  \label{fig:1}
  \vspace{-10pt}
\end{figure}

The conventional end-to-end audiovisual pipeline comprises discrete stages: content uploading, transcoding, content delivery \& network scheduling(pan-CDN resource selection\cite{qiao2025pira} and multipath transport\cite{hu2025cross}), and playback(client-side preloading and download scheduling\cite{liu2025deload})(as shown in \textcolor{blue}{Figure \ref{fig:1}}). The transcoding phase, in particular, involves strategic video processing to generate a bitrate ladder of variants, which is a structured group of video representations as the input of personalized streaming playback on the client-side.

Confronted with daily processing of UGC (User-Generated Content) reaching hundreds of millions (e.g., TikTok/Douyin), short video platforms face severe computational and transmission resource constraints. The challenges in transcoding are manifold: \textbf{i)} How to maximize user experience under limited ladder numbers? In practice, Most videos often offer only a dozen ladder numbers due to ROI considerations.
\textbf{ii)} How to integrate diverse user demands for video quality and playback fluency into bitrate ladder decisions? For instance, as shown in \textcolor{blue}{Figure \ref{fig:1}}, iPhone-7 devices require low-bitrate streams under poor network conditions while iPhone-16 devices demand high-quality streams at high network speeds to satisfy visual quality expectations.

The video encoding optimization problem fundamentally involves the strategic selection of compression parameters (e.g., codec standards, encoding preset,  resolutions, and Constant Rate Factor) to achieve optimal rate-distortion tradeoffs. Both industry and academia have proposed solutions, including: 1) Fixed-configuration encoding\textcolor{blue}{\cite{Apple2023,Google2022}}, which apply uniform parameters regardless of content. 2) Adaptive encoding techniques\textcolor{blue}{\cite{Netflix2015,Netflix2016,de2016complexity,IoannisNetflix2016,chen2018optimized}} such as Content-Aware Encoding (CAE), Per-Title/Per-Scene/Per-Shot/Per-Chunk Encoding, and Context-Aware Encoding\textcolor{blue}{\cite{reznik2018optimal,reznik2019optimal,reznik2023improving}}. 3) Learning-based approaches\textcolor{blue}{\cite{paul2024convex,menon2023jnd,mico2023per}}, which predominantly optimize video proxy metrics (e.g., VMAF). 4) QoE-based approaches\textcolor{blue}{\cite{seufert2015survey,lebreton2021network,lebreton2023quitting,lorenzi2024medusa,bampis2017learning,bampis2018recurrent,bampis2021towards,huang2023optimizing,bentaleb2016sdndash,li2023mamba,deng2024}}.

However, the first three categories of approaches focus solely on content-adaptive video processing, failing to account for authentic user viewing experiences. Meanwhile, the remaining category of methods exhibits limited utilization of QoE principles and cannot address diverse user needs. Current research in this area remains scarce.

\begin{figure*}  
  \centering
  \includegraphics[width=0.9\textwidth]{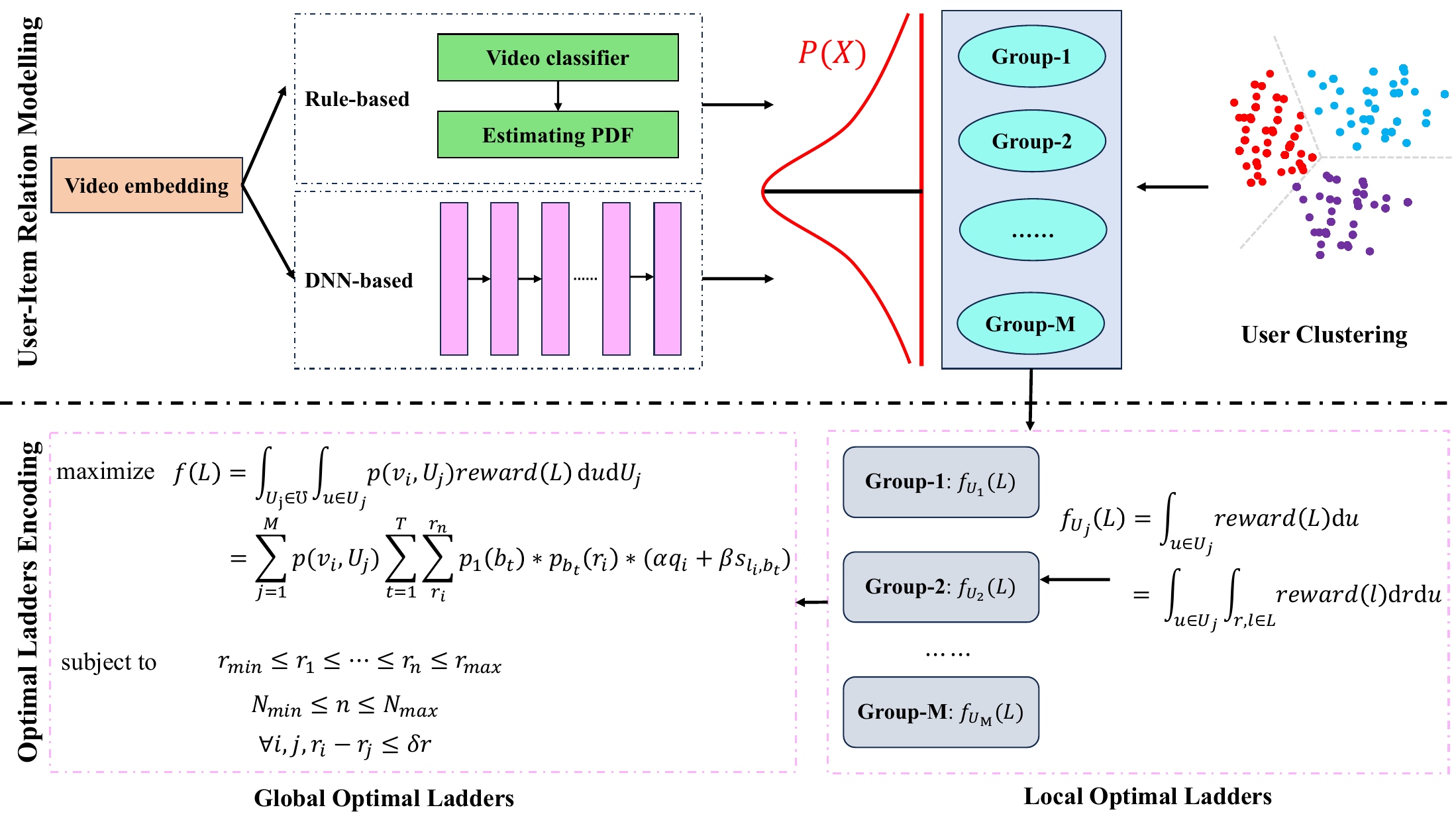}
  \caption{Framework of UIAE.}
  \label{fig:2}
  \vspace{-10pt}
\end{figure*}

To tackle these challenges, this paper proposes a UIAE method tailored for short video scenarios. First, it clusters all consumer groups by similar playback requirements, then establishes user-item relationships using rule-based or DNN-based methods. Next, it develops local optimization models to maximize QoE for sub-populations based on context profiles, and finally forms a globally optimal encoding strategy. Experimental results show the UIAE method exhibits compelling competitiveness compared with state-of-the-art.

\vspace{-10pt}
\section{Proposed Method}
\vspace{-5pt}
In this section, we propose a User-Item Aware Encoding (UIAE) framework tailored for UGC-centric short video scenarios, as illustrated in \textcolor{blue}{Figure \ref{fig:2}}. 

\vspace{-10pt}
\subsection{User-Item Relationship modeling}
\vspace{-5pt}

The goal of the user-item relationship model is to predict the future playback distribution $P(X)$ of a video after upload. We first divide all users into several groups, and then make transcoding decisions based on these groups.

\textbf{User Clustering}. The objective of user clustering is to group users with comparable bitrate ladder clarity/fluency requirements. Generally, users having similar network condition, device hardware, and analogous preferences for video quality and stalling tolerance are anticipated to have similar bitrate demands. Consequently, the following features are primarily considered: 1) device performance indicators, including CPU/GPU model, memory/storage capacity, etc.; 2) historical playback statistics, such as network bandwidth distribution, video quality/stalling statistics, and view completion rates for different bitrate ladders, etc. In this study, the K-means algorithm\textcolor{blue}{\cite{hartigan1979algorithm}} is chosen to do the clustering task. Ultimately, all consumers are partitioned into M groups, i.e., $\mathcal{U}=\{U_1,U_2,...U_M\}$.

\textbf{Rule-based modeling}. This approach first acquires the video's multimodal embedding representation using clip model\textcolor{blue}{\cite{lee2022uniclip,zhou2023non,wang2023video}}, then utilizes the classic multi-head classification model to categorize the video into K categories $\mathcal{V}=\{V_1,V_2,...V_K\}$, where each category represents a specific video type. Subsequently, the playback proportion of each video category in $\{U_1,U_2,...U_M\}$ is calculated from historical playback records. Specifically, the playback distribution $P(X)$ of user-item can be computed by equation \textcolor{blue}{\ref{eq:1}}. For  $\forall v_i^* \in V_i$, the playback probability of $v_i^* $ in group $U_j$ is:

\begin{equation}
\label{eq:1}
p(v_i^*,U_j)=\frac{Play(V_i,U_j)}{Play(V_i,*)}=\frac{\sum_{v_i \in V_i}{Play(v_i,U_j)}}{\sum_{v_i \in V_i}{Play(v_i,*)}}
\end{equation}

where $Play(V_i,U_j)$ denotes the number of playback of all videos in video type $V_i$ within group $U_j$, and $Play(v_i,U_j)$ denotes the number of playback of video $v_i$ within group $U_j$. Notably, $\sum_{j=1}^{M}p(v_i,U_j)=1$, and $p(v_i,U_j) \ge0$.

 \textbf{DNN-based modeling}. In the DNN-based modeling strategy, the model input is $input=\{\mathbf{e_i},\mathbf{a_i}\}$, where $\mathbf{e_i}$ is the multimodal embedding of video ${v_i}$, and $\mathbf{a_i}$ represents the author-side feature information of video ${v_i}$. The model output is an M-dimensional vector, representing the playback probability of M groups, which is implemented using M neurons in the network. The intermediate structure of the model employs multiple simple fully connected layers and normalization layers, and the final activation function is softmax. We use weighted cross-entropy as the loss function :
 \begin{equation}
\mathcal{L}_{WCE}=-\frac{1}{N}\sum_{i=1}^{N}\sum_{j=1}^Mw_j
\cdot y(v_i,U_j)\cdot log(p(v_i,U_j)) 
\end{equation}
where ${w_j}$ is the weight of group ${U_j}$,  ${y(v_i,U_j)} \in \{0, 1\}$ is an indicator function,  ${y(v_i,U_j)} =1$ means that the playback probability of video $v_i$ in group ${U_j}$ is the highest in the real sample. And $p(v_i,U_j)$ represents the predicting playback probability of video $v_i$ in group ${U_j}$ by the DNN model.

\vspace{-10pt}
\subsection{Optimal Ladders Encoding}
\vspace{-5pt}
The optimal multiple bitrate ladder group decision process can be divided into two parts: 1) Local optimization for each subgroup; 2) Global optimization across all consumer users.

\vspace{-10pt}
\subsubsection{\textbf{Local Optimal Ladders}}
\vspace{-5pt}

Given a video $v*$, we define a multiple bitrate ladder group $L$ as a set of ladders $\{l_1,...,l_n\}$ generated by an encoding decision, where all candidate ladder groups form the collection $\mathcal{L}=\{L_1,L_2,...\}$. Each ladder $l_i=(r_i,q_i),i=1,...n$ is characterized by its bitrate $r_i$ and  video quality metric $q_i$ (e,g., VMAF) , and for simplicity, $r_i$ can be used interchangeably with $l_i$ in subsequent descriptions. The bitrates within a ladder group $L$ are ordered in ascending sequence $r_1<r_2<...<r_n$. In this work, all multiple bitrate ladder candidates are generated by adjusting the Constant Rate Factor while fixing other encoder parameters. Then, the optimization problem for selecting the QoE-optimal multiple bitrate ladder group in Group ${U_j}$  can be expressed as follows:
\vspace{-10pt}

\begin{equation}
\label{eq:3}
f_{U_j}(L) = \int_{u \in {U_j}} reward(L){\rm{d}}u
= \int_{u \in {U_j}} \int_{r,l \in {L}}{}reward(l){\rm{d}}r{\rm{d}}u
\end{equation}
Constraint conditions include: 1) bitrate sorting constraint $r_{min} \le r_1\le...\le r_n \le r_{max}$; 2) ladder quantity constraint $N_{min} \le n* \le N_{max}$; 3) bitrate difference constraint between adjacent ladders $\forall i,j, r_i - r_j \le \delta r$. $reward(l)$ in Equation \textcolor{blue}{\ref{eq:3}} quantifies the QoE when playing the ladder $l$.  Here, $U_j$ denotes the set of all users who may play the ladder group $L$.  Typically, we discretize the user set $U_j$ using users’ network bandwidth and playback probability. Specifically,
\begin{equation}
\begin{split}
\label{eq:4}
 \int_{u \in {U_j}} reward(L) {\rm d}u &=\int_{b_t \in {B_j}} reward(L){\rm d}b_t
 \\&=\sum_{t=1}^T\sum_{r_i}^{r_n}p_1(b_t)*p_{b_t}(r_i)*reward(l_i)
\end{split}
\end{equation}
Where $p_1(\cdot)$ represents the network bandwidth probability distribution. In this paper, the network bandwidth is discretized via bucketing. $B_j=\{b_1,b_2,...,b_T\}$ means the bandwidth is divided into $T$ buckets, and $p_{b_t}(r_i)$ denotes the probability of playing the ladder $l_i$ (with bitrate $r_i$) under bandwidth $b_t$. In reality,  $p_{b_t}(*)$ can be derived by fitting statistics from historical user playback data.

The $reward(l_i)$ in Equantion \textcolor{blue}{\ref{eq:4}} quantifies users’ QoE after playing the ladder $l_i$. This paper focuses on two factors most directly affecting user experience: video clarity (e.g., VMAF) and playback fluency (e.g., stalling duration). Notably, stalling duration is based on the two-dimensional distribution of video bitrate and network bandwidth. Let $s_{l_i,b_t}$ represent the stalling duration when playing $l_i$ with network bandwidth $b_t$, thus:
\begin{equation}
reward(l_i)=\alpha_{j}*q_i+\beta_{j}*s_{l_i,b_t}
\end{equation}
$\alpha_j$ and $\beta_j$ are hyperparameters, representing the subjective inclinations of users in Group $U_j$ toward video clarity and playback fluency, respectively.

Finally, Equation \textcolor{blue}{\ref{eq:1}} can be discretized as 
\begin{equation}
f_{U_j}(L)=  \sum_{t=1}^T\sum_{r_i}^{r_n}p_1(b_t)*p_{b_t}(r_i)*(\alpha_j*q_i+\beta_j*s_{l_i,b_t})
\end{equation}

\vspace{-10pt}
\subsubsection{\textbf{Global Optimal Ladders}}
\vspace{-5pt}

Based on the local optimization results for each group $U_j$, the globally optimal multiple bitrate ladder group decision problem for a given video $v*$ can be formally expressed as:

\begin{equation}
\begin{split}
\text{maximize} \quad 
f(L)&= \int_{U_j \subset {\mathcal U}}\int_{u \in {U_j}} p(v_i,U_j)* reward(L){\rm{d}}u{\rm{d}}U_j \\&=\sum_{j=1}^{M}p(v_i,U_j)*\sum_{t=1}^T\sum_{r_i}^{r_n}p_1(b_t)*p_{b_t}(r_i)
\\ & \quad *(\alpha_j*q_i+\beta_j*s_{l_i,b_t})
\\
\\
\text{subject to} \quad  r_{min} & \le r_1\le...\le r_n \le r_{max}
\\
& N_{min} \le n \le N_{max}
\\
& \forall i,j, r_i - r_j \le \delta r
\end{split}
\end{equation}

Clearly, this is a nonlinear constrained optimization problem, and well-established optimization methods such as sequential quadratic programming (SQP) can be selected.

\section{Experiments and Discussion}

\vspace{-5pt}
\subsection{Experimental Setup}
\vspace{-5pt}
\textbf{Video DataSet}. The UIAE framework has already been deployed in Douyin's online transcoding pipeline, handling tens of millions of video transcoding tasks daily to support video playback for hundreds of millions of users. For experimental clarity and simplicity, our validation video dataset was constructed by sampling a subset of videos from the online production environment. This dataset comprised 5 distinct content categories (e.g., Movie/Anime/Gaming/Casual-Vlog/Agriculture), with 10,000 videos in each category. 

\textbf{User-Context.} The user context dataset consists of performance indicators collected from billions of users' online playback data, which are used to generate the probability distributions $p_1(\cdot)$, $p_{b_t}(\cdot)$ and $s_{l_i,b_t}$. During the user clustering stage, $M=4$  (Groups-1 to Group-4) was determined by comparing multiple clustering results.

\textbf{Evaluation metrics}. In this study, the following metrics are used to evaluate the UIAE decision framework: mean \textbf{bitrate}, mean \textbf{VMAF}, \textbf{stalling duration}, \textbf{stalling frequency}, and \textbf{QoE} score. The first four metrics are collected from client-side playback events, while QoE score is computed according to the linear model integrating video quality and stalling penalties. E.g., $QoE=quality(\cdot)-\rho*stalling\_duration(\cdot)-\gamma*stalling\_frequancy(\cdot)$. We empirically configured $\rho=1$ and $\gamma=18$ refer to \textcolor{blue}{\cite{bampis2021towards}}.


    

\begin{table}
  \caption{Comparative analysis of different methods via client-side simulation experiments, with the results are reported as mean values averaged across all metrics. (Bit: Bitrate/Mbps, SD: Stalling-Duration/ms, SF: Stalling-Frequency/\%, HP: Hyperparameter. $\bm{\rho_{1}}=[0.3,0.3,0.3,0.3]$ and $\bm{\rho_{2}}=[0.1,0.4, 0.5,1.5]$ represent uniform and personalized parameter settings, respectively.)}
 \label{tab:1}
  \small
  \setlength{\tabcolsep}{5pt} 
  \begin{tabular}{|c|c|c|c|c|c|c|c|}
    \hline
    N &Methods & HP & QoE & VMAF & Bit & SD & SF\\
    \hline
    \multirow{6}{*}{3}&  $\text{CxAE}_{quality}$ & \multirow{4}{*}{$\bm{\rho_{1}}$} & 56.74 & 93.79 & 1.25 & 25.4 & 0.65 \\
    ~&  $\text{CxAE}_{QoE}       $ & ~ &58.25 & 93.64 & 1.16 & 24.2 & 0.63 \\
    ~&  $\text{UIAE}_{QoE_{rule}}$ & ~ &59.47 & 93.60 & 1.17 & 22.9 & 0.62 \\
    ~&  $\text{UIAE}_{QoE_{dnn}} $ & ~ &\textbf{60.34} & 93.33 & 1.13 & 21.7 & 0.62 \\
    \cline{2-8}
     ~&  $\text{UIAE}_{QoE_{rule}}$ &\multirow{2}{*}{$\bm{\rho_{2}}$} & \textbf{60.13} & 93.24 & 1.18 & 22.2 & 0.61 \\
    ~&  $\text{UIAE}_{QoE_{dnn}} $ & ~ &\textbf{60.86} & 93.13 & 1.12 & 21.2 & 0.60 \\

    \hline
    \multirow{6}{*}{5} & $\text{CxAE}_{quality}$ &\multirow{4}{*}{$\bm{\rho_{1}}$} & 57.13 & 94.56 & 1.55 & 25.5 & 0.66 \\
    ~ & $\text{CxAE}_{QoE}$ & ~ & 58.76 & 94.42 & 1.45 & 24.3 & 0.64 \\
    ~ & $\text{UIAE}_{QoE_{rule}}$ & ~ &59.82 & 94.33 & 1.18 & 23.0 & 0.63 \\
    ~ & $\text{UIAE}_{QoE_{dnn}}$ & ~ & \textbf{60.71} & 94.20 & 1.30 & 21.8 & 0.63 \\
    \cline{2-8}
    ~ & $\text{UIAE}_{QoE_{rule}}$ &\multirow{2}{*}{$\bm{\rho_{2}}$}&\textbf{60.47} & 94.30 & 1.34 & 22.3 & 0.62 \\
    ~ & $\text{UIAE}_{QoE_{dnn}}$ & & \textbf{61.30} & 94.20 & 1.26 & 21.3 & 0.62 \\
    
    \hline
  \end{tabular}
  \vspace{-10pt}
\end{table}

\vspace{-10pt}
\subsection{Results and Analysis}
\vspace{-5pt}

\textbf{Comparison for encoding algorithms}. We compared the results of $\text{CxAE}_{quality}$, $\text{CxAE}_{QoE}$, $\text{UIAE}_{QoE_{rule}}$ and $\text{UIAE}_{QoE_{dnn}}$ on the Video DataSet via client-side simulations. In the encoding phase, both $\text{CxAE}_{quality}$ and $\text{CxAE}_{QoE}$ directly utilizes global context of Group-Overall. $\text{CxAE}_{quality}$ only uses network bandwidth distribution, while $\text{CxAE}_{QoE}$ incorporates both network bandwidth distribution and stalling distribution. The UIAE variants ($\text{UIAE}_{QoE_{rule}}$ and $\text{UIAE}_{QoE_{dnn}}$) employ the local context information of Group-1-4 instead of directly using the global context information of Group-Overall. The experiment compared two cases of ladder numbers: 3 and 5.



\textcolor{blue}{Table \ref{tab:1}} presents the performance of different methods under ladder numbers 3/5 and hyperparameter vectors $\bm{\rho_1}$/$\bm{\rho_2}$. The three QoE-based methods outperform the quality-oriented  $\text{CxAE}_{quality}$. For instance, when N=5 and HP=$\bm{\rho_1}$, $\text{UIAE}_{QoE_{dnn}}$ achieves a QoE of 60.71, surpassing $\text{CxAE}_{quality}$'s 57.13. Meanwhile, its stalling-duration of 21.8 ms and stalling-frequency of 0.63\% are also superior to $\text{CxAE}_{quality}$’s 25.5 ms and 0.66\%, thus better aligning with user viewing needs. Among QoE-based methods, two UIAE variants excel further. In the 5-tier configuration, these enhancements persist: QoE improvements of 1.06 and 1.95 are observed, alongside stalling duration reductions of 1.3 ms (-5\%) and 2.5 ms (-10\%).  In summary, both the $\text{UIAE}_{QoE_{dnn}}$ and $\text{UIAE}_{QoE_{rule}}$ methods demonstrate superior QoS and QoE performance in client-side simulations, particularly excelling in fluency metrics (stalling duration/frequency), owing to the granular integration of user-contextual data into the multiple bitrate ladder optimization objective.

\begin{figure}
  \centering
  \includegraphics[width=\columnwidth]{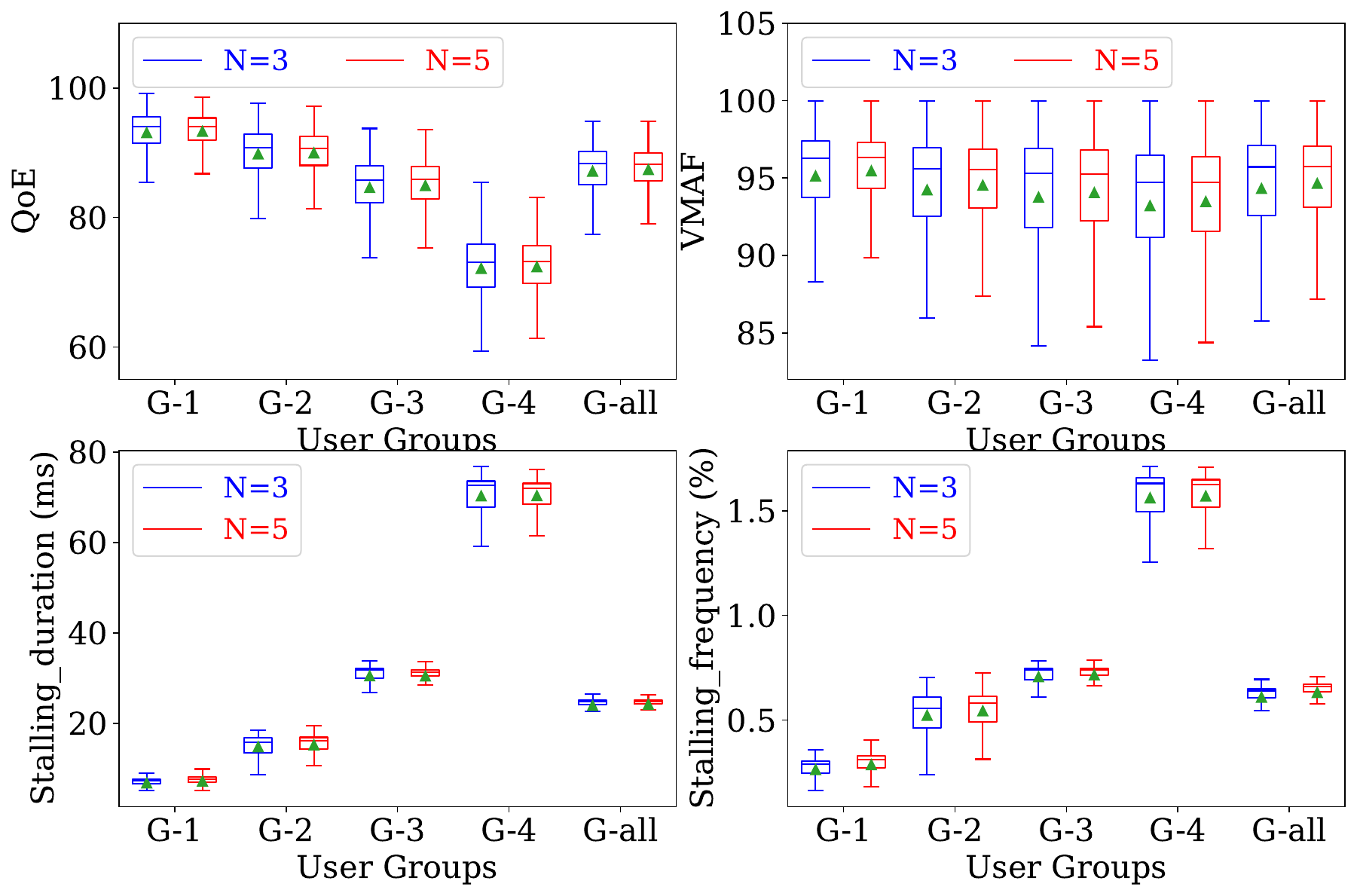}

  \vspace{-10pt}
  \caption{Comparative analysis on $\text{CxAE}_{QoE}$ with heterogeneous user contexts. G-1 to G-all denote Group-1 to Group-overall. The hyperparameter $\rho=-\frac{\beta_j}{\alpha_j}=0.3$.}
  
  \label{fig:3}
  \vspace{-10pt}
\end{figure}

\textbf{Importance of user context.} This subsection analyzes the impact of user context variations on encoding outcomes through controlled experiments with the $\text{CxAE}_{QoE}$ method across all user groups (Network bandwidth and stalling events deteriorate progressively from Group-1 to Group-4). The average simulation results are shown in \textcolor{blue}{Figure \ref{fig:3}}. Group 4, characterized by poorer user context profiles, exhibit degraded performance across all metrics. Specifically, Groups 1-3 achieved higher VMAF scores, significantly lower stalling metrics, resulting in a significantly better QoE distribution. These results underscore the critical role of user context: uniform global parameters fail to accommodate heterogeneous network bandwidth and stalling constraints.

\textbf{Analysis Parameters for Groups.} Based on the performance of Group-1 to Group-4, we slightly adjusted the hyperparameters vector $\bm{\rho_{2}}$, with specific settings as follows: $-\frac{\beta_1}{\alpha_1}=0.1$, $-\frac{\beta_2}{\alpha_2}=0.4$, $-\frac{\beta_3}{\alpha_3}=0.5$, $-\frac{\beta_4}{\alpha_4}=1.5$. The comparison results against the baseline uniform setting $\bm{\rho_1}$ are shown in \textcolor{blue}{Table \ref{tab:1}}. Both $\text{UIAE}_{QoE_{dnn}}$ and $\text{UIAE}_{QoE_{rule}}$ achieve superior QoE through cluster-specific parameterization. Specifically, when the ladder numbers = 3, the QoE of $\text{UIAE}_{QoE_{dnn}}$ and $\text{UIAE}_{QoE_{rule}}$ increases by 0.52 (+0.86\%) and 0.66 (+1.11\%), respectively. This personalized optimization enables the model to better align bitrate decisions with subgroup-specific constraints.

\vspace{-10pt}
\section{Conclusion}
\vspace{-5pt}
We present UIAE, a novel multiple bitrate ladder optimization framework. It establishes user-item relationships via either rule-based or DNN-based methods and then formalizes the multi-bitrate ladder decision problem as a global QoE maximization problem through hierarchical model integration. Experimental results have demonstrated the superiority of UIAE. Future work will concentrate on personalized preference modeling using multi-modal behavior data and edge-compatible lightweight adaptation, progressing towards truly context-aware streaming systems for heterogeneous UGC ecosystems.

\clearpage
\vfill\pagebreak

\bibliographystyle{IEEEbib}

\bibliography{submit}

\end{document}